\documentclass[conference]{IEEEtran} % IEEE HPEC
\IEEEoverridecommandlockouts
\usepackage{cite}
\usepackage{amsmath,amssymb,amsfonts}
\usepackage{algorithmic}
\usepackage{graphicx}
\usepackage{textcomp}
\usepackage{xcolor}
\def\BibTeX{{\rm B\kern-.05em{\sc i\kern-.025em b}\kern-.08em
    T\kern-.1667em\lower.7ex\hbox{E}\kern-.125emX}}
\begin{document}

\title{Focusing and Calibration of Large Scale Network Sensors using GraphBLAS Anonymized Hypersparse Matrices 
\thanks{This material is based upon work supported by the Under Secretary of Defense for Research and Engineering under Air Force Contract No. FA8702-15-D-0001. Any opinions, findings, conclusions or recommendations expressed in this material are those of the author(s) and do not necessarily reflect the views of the Under Secretary of Defense for Research and Engineering. Research was also sponsored by the United States Air Force Research Laboratory and the Department of the Air Force Artificial Intelligence Accelerator and was accomplished under Cooperative Agreement Number FA8750-19-2-1000. The views and conclusions contained in this document are those of the authors and should not be interpreted as representing the official policies, either expressed or implied, of the Department of the Air Force or the U.S. Government. The U.S. Government is authorized to reproduce and distribute reprints for Government purposes notwithstanding any copyright notation herein.}
}

\author{\IEEEauthorblockN{Jeremy Kepner$^1$, Michael Jones$^1$, Phil Dykstra$^2$, Chansup Byun$^1$, Timothy Davis$^3$ ,  Hayden Jananthan$^1$,  \\ William Arcand$^1$, David Bestor$^1$, William Bergeron$^1$, Vijay Gadepally$^1$,  Micheal Houle$^1$, \\ Matthew Hubbell$^1$, Anna Klein$^1$, Lauren Milechin$^1$, Guillermo Morales$^1$, Julie Mullen$^1$, \\ Ritesh Patel$^1$, Alex Pentland$^1$, Sandeep Pisharody$^1$, Andrew Prout$^1$,  Albert Reuther$^1$, Antonio Rosa$^1$, \\  Siddharth Samsi$^1$, Tyler Trigg$^1$, Charles Yee$^1$, Peter Michaleas$^1$
\\
\IEEEauthorblockA{$^1$MIT,  $^2$HPCMP DREN, $^3$Texas A\&M,
}}}
\maketitle

\begin{abstract}
Defending community-owned cyber space requires community-based efforts.  Large-scale network observations that uphold the highest regard for privacy are key to protecting our shared cyberspace.  Deployment of the necessary network sensors requires careful sensor placement, focusing, and calibration with significant volumes of network observations.  This paper demonstrates novel focusing and calibration procedures on a multi-billion packet dataset using high-performance GraphBLAS anonymized hypersparse matrices.  The run-time performance on a real-world data set confirms previously observed real-time processing rates for high-bandwidth links while achieving  significant data compression.   The output of the analysis demonstrates the effectiveness of these procedures at focusing the traffic matrix and revealing the underlying stable heavy-tail statistical distributions that are necessary for anomaly detection.  A simple model of the corresponding probability of detection ($p_{\rm d}$) and probability of false alarm ($p_{\rm fa}$) for these distributions highlights the criticality of network sensor focusing and calibration.  Once a sensor is properly focused and calibrated it is then in a position to carry out two of the central tenets of good cybersecurity: (1) continuous observation of the network and (2) minimizing unbrokered network connections. 
\end{abstract}

\begin{IEEEkeywords}
Internet defense, packet capture, streaming graphs, hypersparse matrices
\end{IEEEkeywords}

\section{Introduction}

\begin{figure}
\center{\includegraphics[width=1.0\columnwidth]{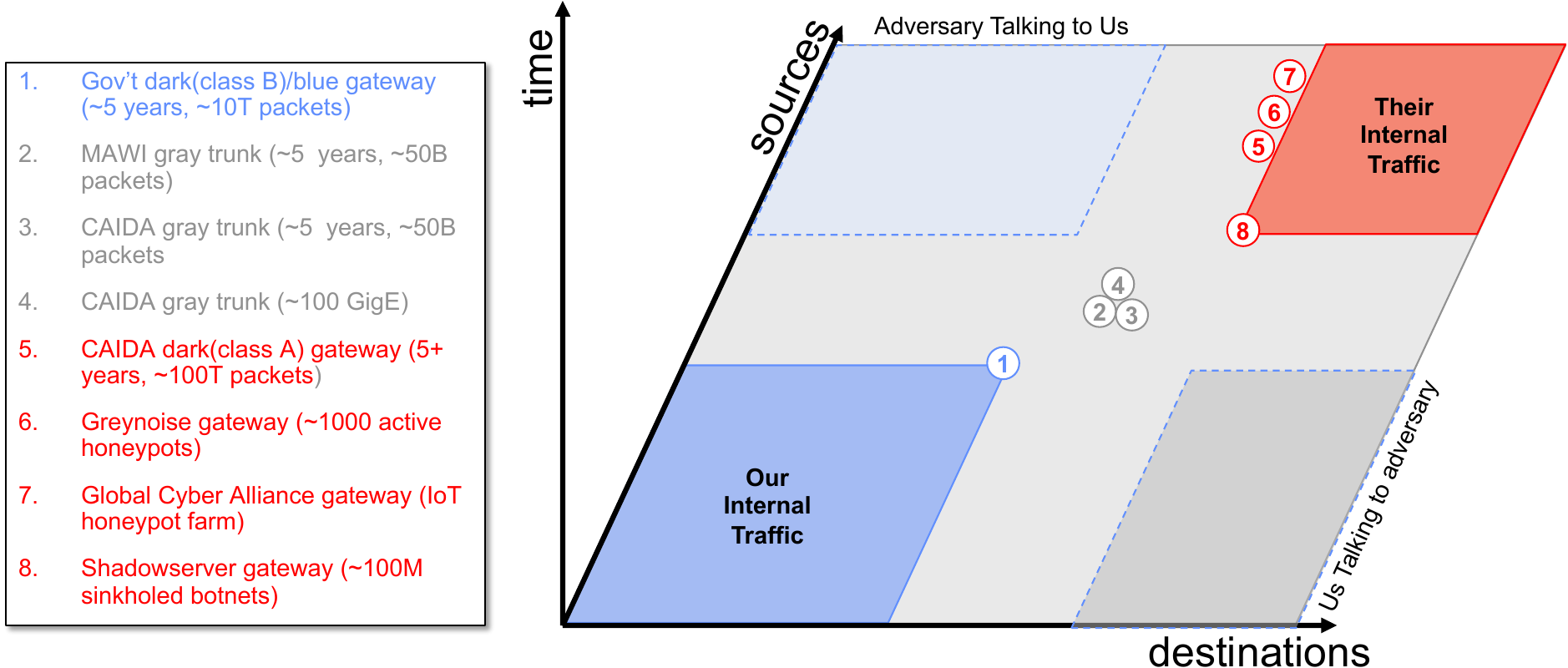}}
      	\caption{{\bf Network Sensor Placement.}  Traffic matrix view of the Internet depicting selected network sensors and their notional proximity to various types of network traffic \cite{CAIDA2019, CAIDA2022, GCA2022, Greynoise2022, MAWI2020, Shadowserver2022, kepner2020multi}.   The location of a sensor determines the expected distributions of  the observations.  Figure adapted from \cite{kepner2021zero}.}
      	\label{fig:observatories}
\end{figure}

Our community-owned cyber space (industry, government, and academia) requires community-based protections involving significant data sharing  \cite{atkins2021improvised, atkins2021cooperation, demchak2021achieving, weed2022beyond, weed2023beyond}.  Large-scale network observations with the highest regard for privacy are key  \cite{kepner2021zero, pisharody2021realizing, pentland2022building}.
In the cyber domain, observatories and outposts have been constructed to gather data on  Internet traffic and provide a starting point for exploring community-based approaches \cite{CAIDA2019, CAIDA2022, GCA2022, Greynoise2022, MAWI2020, Shadowserver2022, kepner2020multi}  (see Figure~\ref{fig:observatories}).
Effective deployment of a sensor requires careful sensor placement, focusing, and calibration with significant volumes of  observations.  This is particularly important for distributed sensors \cite{bychkovskiy2003collaborative}.

The data volumes, processing requirements, and privacy concerns of analyzing a significant fraction of the Internet have been prohibitive.  The North American Internet generates billions of non-video Internet packets each second \cite{Cisco2017, Cisco2018-2023}.   The GraphBLAS standard  provides significant performance and compression capabilities which improve the feasibility of analyzing these volumes of data \cite{kepner16mathematical, buluc17design,yang2018implementing, kepner2018mathematics, davis2019algorithm, mattson2019lagraph, cailliau2019redisgraph, davis2019write, aznaveh2020parallel, brock2021introduction, pelletier2021graphblas, jones2022graphblas, trigg2022hypersparse}.   Specifically, the GraphBLAS is ideally suited for both constructing and analyzing anonymized hypersparse traffic matrices.  Prior   GraphBLAS work has demonstrated rates of 200 billion packets per second (pps) on a supercomputer \cite{kepner2021vertical}, while achieving compressions of 1 bit per packet \cite{kepner2020multi}, and enabling the analysis of the largest publicly available historical archives with over 40 trillion packets \cite{kepner2021spatial}. 
% Analysis of anonymized hypersparse traffic matrices from a variety of sources has revealed power-law distributions \cite{kepner19hypersparse, kepner2022new}, novel scaling relations \cite{kepner2020multi, kepner2021spatial, kepner2022temporal}, and inspired new models of network traffic \cite{devlin2021hybrid}.

GraphBLAS anonymized hypersparse traffic matrices represent one set of design choices for analyzing network traffic.  Specifically, the use case requiring some data on all packets (no down-sampling), high performance, high compression,  matrix-based analysis, anonymization, and open standards.  There are a wide range of alternative graph/network analysis technologies and many good implementations  achieve performance close to the limits of the underlying computing hardware  \cite{tumeo2010efficient, kumar2018ibm, ezick2019combining, gera2020traversing, azad2020evaluation, du2021interactive, acer2021exagraph, blanco2021delayed, ahmed2021online, azad2021combinatorial, koutra2021power}.  Likewise, there are many network analysis tools that focus on providing a rich interface to the full diversity of data found in network traffic \cite{hofstede2014flow, sommer2003bro, lucente2008pmacct}.  Each of these technologies has appropriate use cases in the broad field of Internet traffic analysis.

The ability to rapidly process enormous volumes of anonymized network traffic enables detailed analysis, which is dependent upon proper focusing (understanding precisely where the sensor is) and calibration (understanding the expected statistical distributions of data).  This work explores novel focusing and calibration procedures on a multi-billion packet dataset using high-performance GraphBLAS anonymized hypersparse matrices.

The outline of the rest of the paper is as follows.  First, some basic network quantities  are defined in terms of traffic matrices.   Next the focusing and calibration procedures are described and demonstrated on a gateway dataset revealing the underlying  heavy-tail statistical distributions.  A simple model of the corresponding probability of detection ($p_{\rm d}$) and probability of false alarm ($p_{\rm fa}$) for these distributions is presented that highlights the criticality of network sensor focusing and calibration. The paper concludes with a summary and discussion of future work.

\section{Traffic Matrices and Network Quantities}

Internet data  must be handled with care.  The Center for Applied Internet Data Analysis (CAIDA) based at the University of California's San Diego Supercomputer Center has pioneered  trusted data sharing best practices that combine anonymizing source and destinations using CryptoPAN \cite{fan2004prefix} with data sharing agreements. These data sharing best practices include the following principles  \cite{kepner2021zero} 
\begin{itemize}
\item Data is made available in curated repositories
\item Using standard anonymization methods where needed: hashing, sampling, and/or simulation
\item Registration with a repository and demonstration of legitimate research need
\item Recipients legally agree to neither repost a corpus nor deanonymize data
\item Recipients can publish analysis and data examples necessary to review research
\item Recipients agree to cite the repository and provide publications back to the repository
\item Repositories can curate enriched products developed by researchers
\end{itemize}

\begin{figure}
\center{\includegraphics[width=1.0\columnwidth]{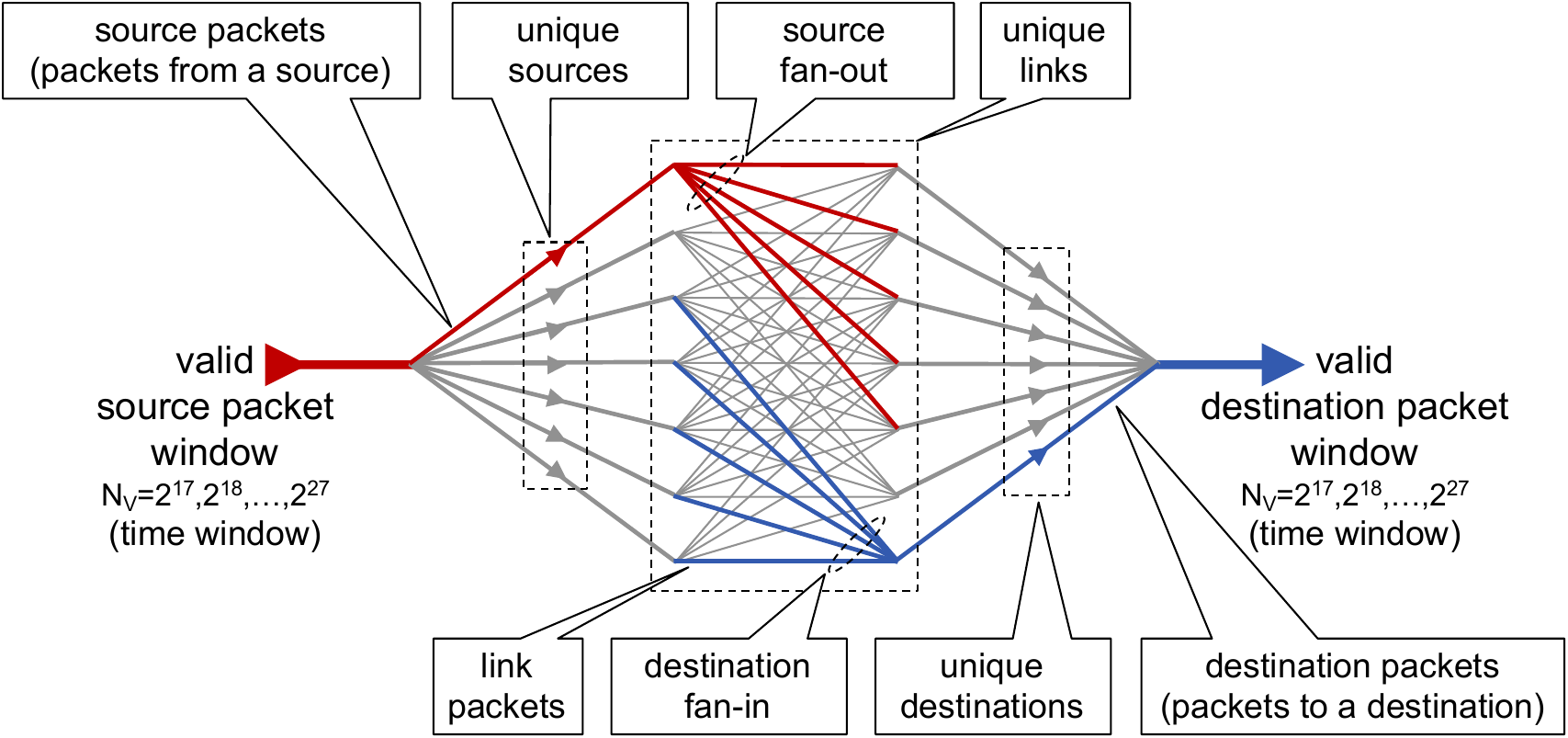}}
      	\caption{{\bf Streaming Network Traffic Quantities.} Internet traffic streams of $N_V$ valid packets are divided into a variety of quantities for analysis: source packets, source fan-out, unique source-destination pair packets (or links), destination fan-in, and destination packets.  Figure adapted from \cite{kepner19hypersparse}.}
      	\label{fig:NetworkDistribution}
\end{figure}

Network traffic data can be viewed as a traffic matrix where each row is a source and each column is a destination (see Figure~\ref{fig:observatories}). A primary benefit of constructing anonymized hypersparse traffic matrices with the GraphBLAS is the efficient computation of a wide range of network quantities via matrix mathematics.  Figure~\ref{fig:NetworkDistribution} illustrates essential quantities found in all streaming dynamic networks. These quantities are all computable from anonymized traffic matrices created from the source and destinations found in Internet packet headers.

\begin{table}
\caption{Network Quantities from Traffic Matrices}
\vspace{-0.25cm}
Formulas for computing network quantities from  traffic matrix ${\bf A}_t$ at time $t$ in both summation and matrix notation. ${\bf 1}$ is a column vector of all 1's, $^{\sf T}$  is the transpose operation, and $|~|_0$ is the zero-norm that sets each nonzero value of its argument to 1\cite{karvanen2003measuring}.  These formulas are unaffected by matrix permutations and will work on anonymized data.  Table adapted from \cite{kepner2020multi}.
\begin{center}
\begin{tabular}{p{1.45in}p{0.9in}p{0.6in}}
\hline
{\bf Aggregate} & {\bf ~~~~Summation} & {\bf ~Matrix} \\
{\bf Property} & {\bf ~~~~~~Notation} & {\bf Notation} \\
\hline
Valid packets $N_V$ & $~~\sum_i ~ \sum_j ~ {\bf A}_t(i,j)$ & $~{\bf 1}^{\sf T} {\bf A}_t {\bf 1}$ \\
Unique links & $~~\sum_i ~ \sum_j |{\bf A}_t(i,j)|_0$  & ${\bf 1}^{\sf T}|{\bf A}_t|_0 {\bf 1}$ \\
Link packets from $i$ to $j$ & $~~~~~~~~~~~~~~{\bf A}_t(i,j)$ & ~~~$~{\bf A}_t$ \\
Max link packets ($d_{\rm max}$) & $~~~~~\max_{ij}{\bf A}_t(i,j)$ & $\max({\bf A}_t)$ \\
\hline
Unique sources & $~\sum_i |\sum_j ~ {\bf A}_t(i,j)|_0$  & ${\bf 1}^{\sf T}|{\bf A}_t {\bf 1}|_0$ \\
Packets from source $i$ & $~~~~~~~\sum_j ~ {\bf A}_t(i,j)$ & ~~$~~{\bf A}_t  {\bf 1}$ \\
Max source packets ($d_{\rm max}$)  & $ \max_i \sum_j ~ {\bf A}_t(i,j)$ & $\max({\bf A}_t {\bf 1})$ \\
Source fan-out from $i$ & $~~~~~~~~~~\sum_j |{\bf A}_t(i,j)|_0$  & ~~~$|{\bf A}_t|_0 {\bf 1}$ \\
Max source fan-out ($d_{\rm max}$) & $ \max_i \sum_j |{\bf A}_t(i,j)|_0$  & $\max(|{\bf A}_t|_0 {\bf 1})$ \\
\hline
Unique destinations & $~\sum_j |\sum_i ~ {\bf A}_t(i,j)|_0$ & $|{\bf 1}^{\sf T} {\bf A}_t|_0 {\bf 1}$ \\
Destination packets to $j$ & $~~~~~~~\sum_i ~ {\bf A}_t(i,j)$ & ${\bf 1}^{\sf T}|{\bf A}_t|_0$ \\
Max destination packets ($d_{\rm max}$) & $ \max_j \sum_i ~ {\bf A}_t(i,j)$ & $\max({\bf 1}^{\sf T}|{\bf A}_t|_0)$ \\
Destination fan-in to $j$ & $~~~~~~~~~~\sum_i |{\bf A}_t(i,j)|_0$ & ${\bf 1}^{\sf T}~{\bf A}_t$ \\
Max destination fan-in ($d_{\rm max}$) & $ \max_j \sum_i |{\bf A}_t(i,j)|_0$ & $\max({\bf 1}^{\sf T}~{\bf A}_t)$ \\
\hline
\end{tabular}
\end{center}
\label{tab:Aggregates}
\end{table}%

The network quantities depicted in Figure~\ref{fig:NetworkDistribution} are computable from anonymized origin-destination traffic  matrices that are widely used to represent network traffic \cite{soule2004identify, zhang2005estimating, mucha2010community, tune2013internet}.  It is common to filter the packets down to a valid set for  any particular analysis.   Such filters may limit particular sources, destinations, protocols, and time windows. To reduce statistical fluctuations, the streaming data should be partitioned so that for any chosen time window all data sets have the same number of valid packets \cite{kepner19streaming}.  At a given time $t$, $N_V$ consecutive valid packets are aggregated from the traffic into a hypersparse matrix ${\bf A}_t$, where ${\bf A}_t(i,j)$ is the number of valid packets between the source $i$ and destination $j$. The sum of all the entries in ${\bf A}_t$ is equal to $N_V$
$$
    \sum_{i,j} {\bf A}_t(i,j) = N_V
$$
Constant packet, variable time samples simplify the statistical analysis of the heavy-tail distributions commonly found in network traffic quantities \cite{kepner19hypersparse, nair2020fundamentals, kepner2022new}.  All the network quantities depicted in Figure~\ref{fig:NetworkDistribution} can be readily computed from ${\bf A}_t$ using the formulas listed in Table~\ref{tab:Aggregates}.  Because matrix operations are generally invariant to permutation (reordering of the rows and columns), these quantities can readily be computed from anonymized data.  Furthermore, the anonymized data can be analyzed by source and destination subranges (subsets when anonymized)  using simple matrix multiplication.  For a given subrange represented by an anonymized hypersparse diagonal matrix ${\bf A}_r$, where ${\bf A}_r(i,i) = 1$ implies  source/destination $i$ is in the range, the traffic within the subrange can be computed via: ${\bf A}_r {\bf A}_t  {\bf A}_r$. Likewise, for additional privacy guarantees that can be implemented at the  edge, the same method can be used to exclude a range of data from the traffic matrix
$$
     {\bf A}_t - {\bf A}_r {\bf A}_t  {\bf A}_r
$$ 

\begin{figure}
\center{\includegraphics[width=1.0\columnwidth]{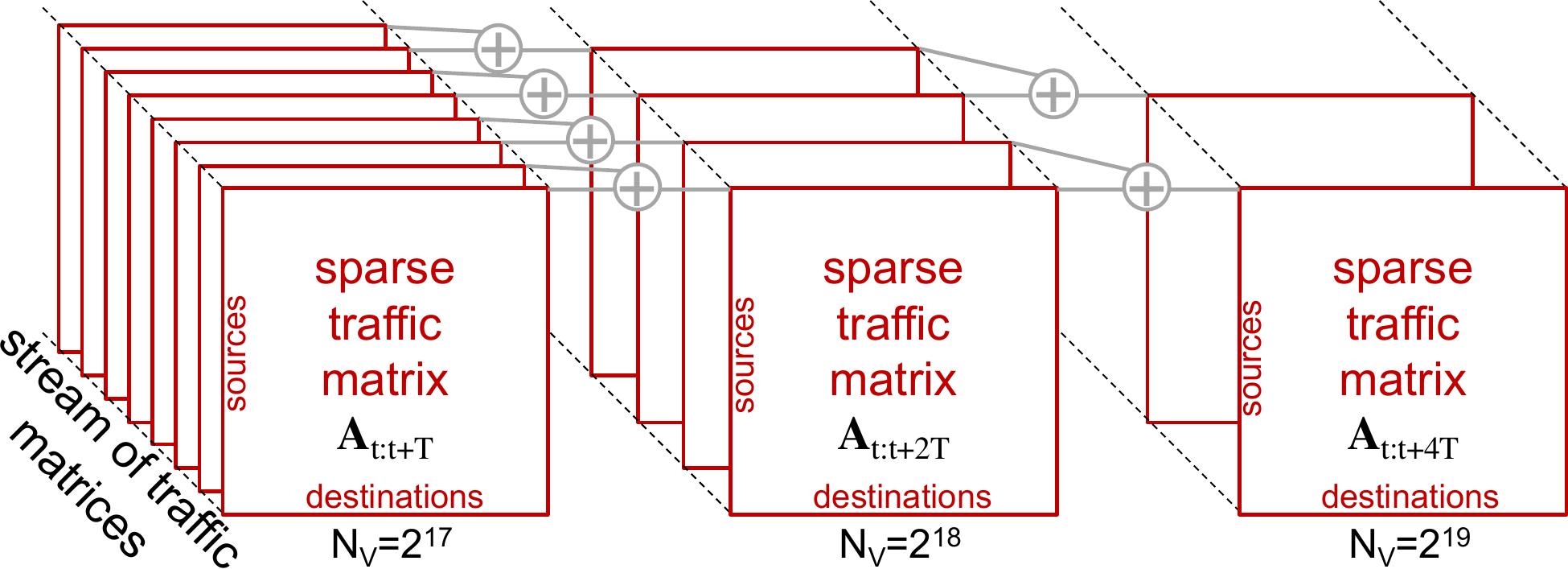}}
      	\caption{{\bf Multi-temporal streaming traffic matrices.} Efficient computation of network quantities on multiple time scales can be achieved by hierarchically aggregating data in different time windows.  Figure adapted from \cite{kepner2020multi}.}
      	\label{fig:MultiTemporalMatrix}
\end{figure}

Efficient computation of network quantities on multiple time scales can be achieved by hierarchically aggregating data in different time windows \cite{kepner19streaming}.  Figure~\ref{fig:MultiTemporalMatrix} illustrates a binary aggregation of  different streaming traffic matrices.   Computing each quantity at each hierarchy level eliminates redundant computations that would be performed if each packet window was computed separately.  Hierarchy also ensures that most computations are performed on smaller matrices residing in faster memory.  Correlations among the matrices mean  that adding two matrices each with $N_V$ entries results in a matrix with fewer than $2N_V$ entries, reducing the relative number of operations as the matrices grow.

One of the important capabilities of the SuiteSparse GraphBLAS library is direct support of hypersparse matrices where the number of nonzero entries is significantly less than either dimensions of the matrix \cite{bulucc2009parallel}.  If the packet source and destination identifiers are drawn from a large numeric range, such as those used in the Internet protocol, then a hypersparse representation of ${\bf A}_t$ eliminates the need to keep track of additional indices and can significantly accelerate the computations \cite{kepner202075}.

\section{Network Sensor Focusing \& Calibration}

The process for rapid construction and analysis of GraphBLAS hypersparse traffic matrices is described in \cite{jones2022graphblas, trigg2022hypersparse}  and briefly summarized here.  The first step in the GraphBLAS hypersparse traffic matrix pipeline is to capture a packet, discard the data payload, and extract the source and destination Internet Protocol (IP) addresses.   For the purposes of the current testing, only IPv4 packets are used which are stored as 32 bit unsigned integers.  Collections of $N_V = 2^{17}$ consecutive packets can each be anonymized with CryptoPAN directly or using a CryptoPAN generated anonymization table.  The resulting anonymized source and destination IPs are then used to construct a $2^{32}{\times}2^{32}$ hypersparse GraphBLAS matrix.  64 consecutive hypersparse GraphBLAS matrices are each serialized in compressed sparse rows (CSR) format with ZSTD \cite{collet2018zstandard} compression and saved to a UNIX TAR file.

\begin{figure}
\center{\includegraphics[width=0.85\columnwidth]{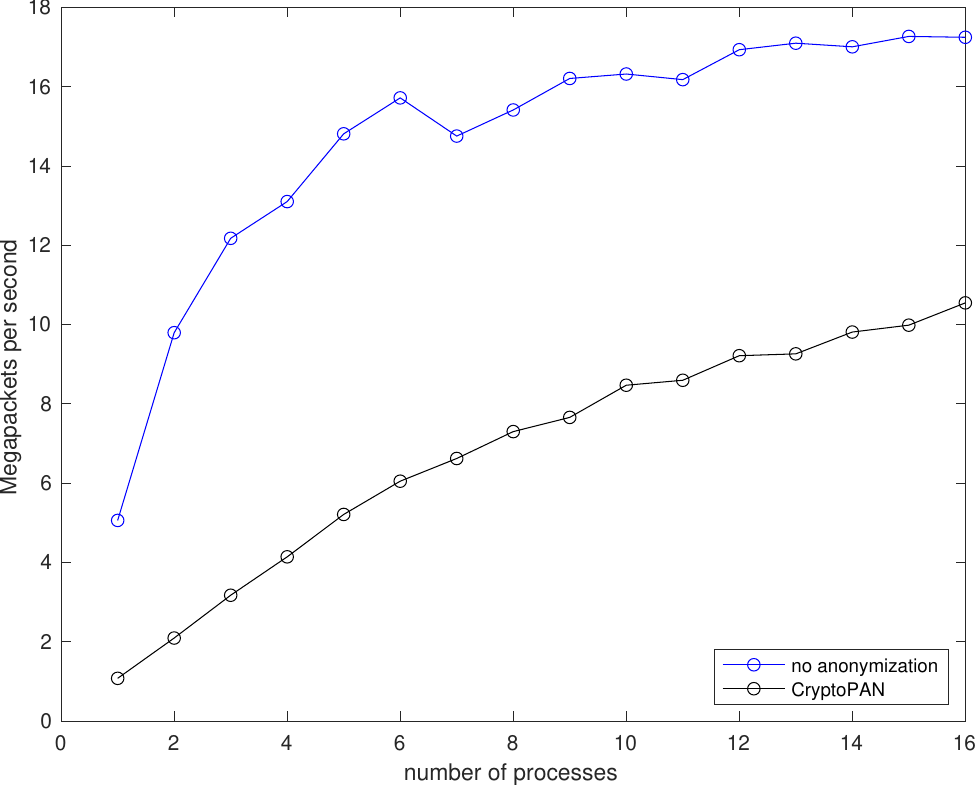}}
      	\caption{{\bf Parallel Performance.} Single-node parallel performance of the hypersparse GraphBLAS construction program running on multiple distinct PCAP files both with and without direct CryptoPAN anonymization.}
      	\label{fig:PacketRate}
\end{figure}

This procedure \cite{jones2022graphblas} was applied to a 246 GB packet capture (PCAP) dataset collected at a gateway for  several  minutes consisting of $2^{31} = 2,147,483,648$ packets.  The hypersparse GraphBLAS matrices were constructed by running the aforementioned procedure on a 32 core 2.3 GHz Xeon 6314 server with 128 GB of RAM and a 7.8 TB NVMe storage device.  The resulting 256 compressed GraphBLAS TAR files, each encompassing $64{\times}2^{17} = $ 8,388,608 packets, had an average file size of 7.4MB corresponding to $<1$ byte per packet.  These files were further analyzed \cite{trigg2022hypersparse} to compute all the network quantities in Table~\ref{tab:Aggregates} for packet windows $N_V=2^{17},\ldots,2^{27}$ over a $4{\times}4$ set of sub-ranges resulting in a single 4.7MB file corresponding to $<0.01$ bits per packet.  Both of these steps highlight the significant compression (100x and 5000x) of these GraphBLAS analysis procedures on a real-world data set. The single-core single-thread GraphBLAS processing time for this dataset (including IO) was $\approx 3x$ faster than real-time for this 6.8 Gigabit/second network gateway.  Figure~\ref{fig:PacketRate} shows the single-node parallel performance of the hypersparse GraphBLAS construction program running on multiple distinct PCAP files both with and without direct CryptoPAN anonymization.

\begin{figure}
\center{\includegraphics[width=0.65\columnwidth]{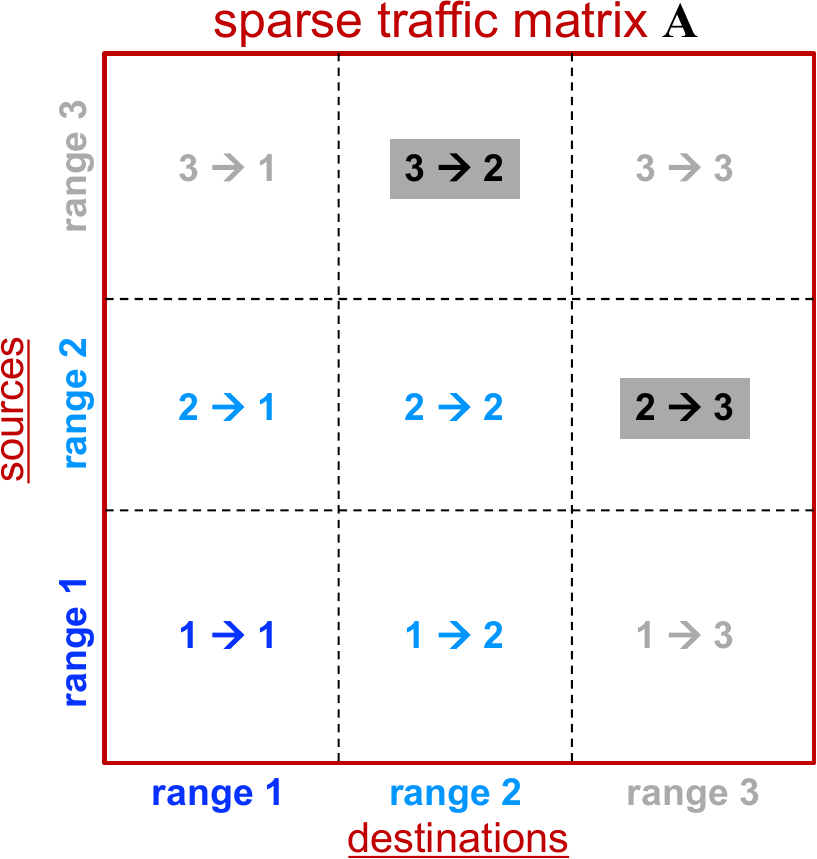}}
      	\caption{{\bf Network Sensor Traffic Matrix Ranges.} The traffic matrix for a given sensor can be divided into ranges separating different categories of traffic.  If range 1 is internal traffic, range 2 is at the gateway, and range 3 is external traffic, then a gateway sensor would be expected to see primarily range 2 to range 3 and range 3 to range 2 traffic.}
      	\label{fig:GatewayTrafficMatrix}
\end{figure}

The focusing procedure begins with understanding the expected location of the network sensor with respect to the various sub-ranges. Figure~\ref{fig:GatewayTrafficMatrix} illustrates a typical position for a gateway sensor.  The expectation of such a sensor is that it will primarily see traffic outbound and inbound from the rest of the world.  Our test data set is processed using 4 subranges corresponding to internal nonroutable ($2^{24}+2^{}+2^{20}+2^{16} =$ 17,891,328 IP addresses), internal bogon ($2^{22} + 2^{17} + 2^{16} + 6{\times}2^{8} =$ 4,392,448 IP addresses), assigned to the gateway ($\approx 3.6$million IP addresses), and other addresses corresponding to the rest of the Internet ($\approx 4.2$ billion IP addresses).  A focused traffic matrix will have an observed statistical distribution of packets that is consistent with the location of the sensor.

\begin{figure}
\center{\includegraphics[width=0.85\columnwidth]{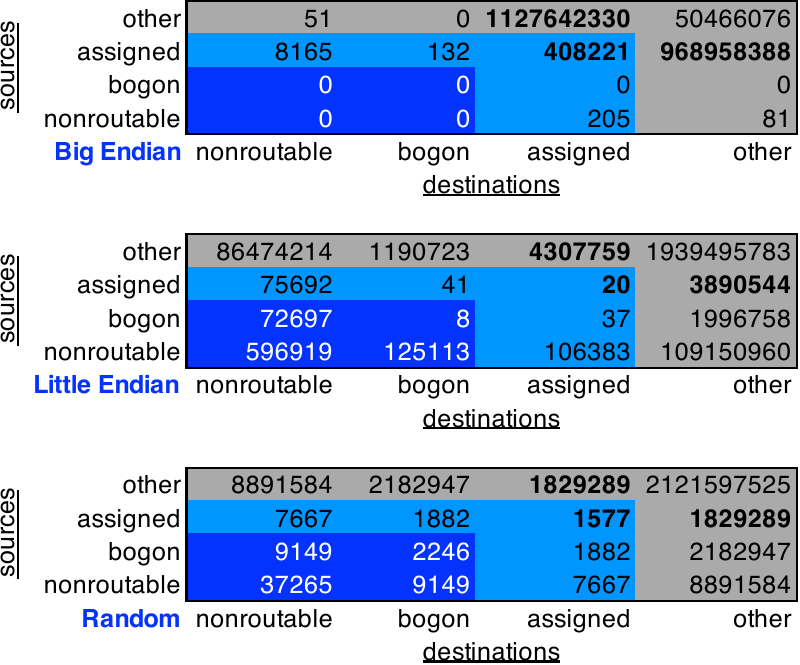}}
      	\caption{{\bf Traffic Matrix Focus Table.} A focused traffic matrix will have an observed statistical distribution of packets that is consistent with the location of the sensor. This can be illustrated with determining the ``endianness'' of the data.  It is possible to estimate the expected distribution of these packets if they are randomly distributed (bottom).  Incorrect endianness will defocus the data and make it look more like a random distribution.  The big-endian (top) and little-endian (middle) representations of the observed data indicate that big-endian is less random and little-endian is more random, thus the underlying data is big-endian given that the sensor is expected to be at a gateway that sees mostly assigned to/from other traffic.}
      	\label{fig:FocusTable}
\end{figure}

The focusing procedure can be illustrated in Figure~\ref{fig:FocusTable} by determining the ``endianness'' of the data, which may not be apparent in extremely large samples.  Generally, network traffic is big-endian in flight, but can be either big-endian or little-endian when stored for processing.  Incorrect endianness will defocus the data and make it look more like a random distribution.  The big-endian and little-endian representations of the observed data indicate that big-endian is less random and little-endian is more random, thus the underlying data is big-endian given that the sensor is expected to be at a gateway that sees mostly assigned to/from other traffic.  The above focusing procedure can be used to correct/check for any distortions in the data and is repeated until  the expected distribution is consistent with  the known location of the sensor.  Likewise, the procedure can be used to quickly determine if a sensor is in a location different from what is expected, which can easily happen when a large number of network sensors are deployed.  Once a sensor is focused, the appearance of traffic in sub-ranges where none is expected can be a simple way to detect anomalies.  For example, an emerging best-practice is to limit unbrokered network connections \cite{arsenault2021hybrid}, which can imply that all traffic must pass through a defined gateway and no other traffic should be observed.

\begin{figure}
\center{\includegraphics[width=0.85\columnwidth]{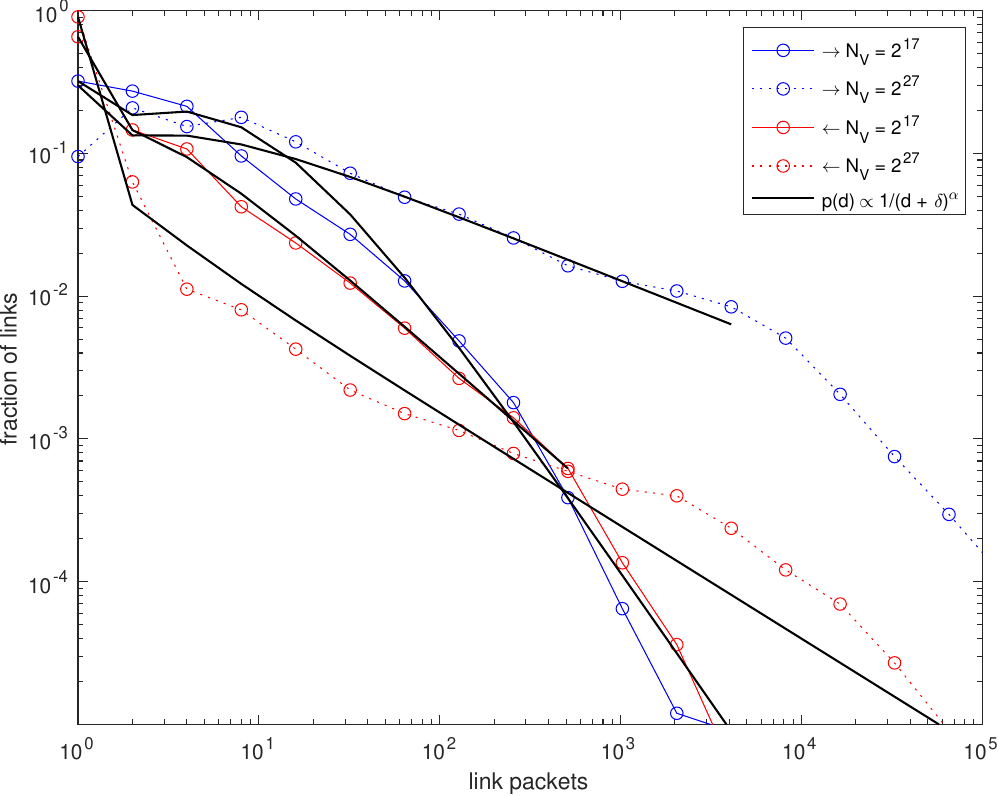}}
      	\caption{{\bf Observed Heavy-Tail Distributions.} Fraction of observed links that have a specified number of packets (d)  showing the strong dependence on the size of the packet window ($N_V$) and direction of the traffic:  assigned-to-other ($\rightarrow$) and other-to-assigned ($\leftarrow$).  Solid black lines show best fit Zipf-Mandelbrot distributions as a function of $d$.}
      	\label{fig:CalibrationPlot}
\end{figure}

  Once the network sensor has been focused, the data in each of the sub-ranges can be analyzed.  Figure~\ref{fig:CalibrationPlot} shows the distribution of the number of packets going over specific links (source/destination pairs) for different sub-ranges (assigned-to-other and other-to-assigned) and packet window sizes $N_V=2^{17}, 2^{27}$.  These results  show the widely observed heavy-tail distributions that are well modeled by the Zipf-Mandelbrot distribution \cite{kepner19hypersparse, kepner2022new}
$$
  p(d) \propto 1/(d + \delta)^\alpha
$$
where $d$ is the number packets observed traversing a link and $\delta$ and $\alpha$ are best-fit parameters.   Figure~\ref{fig:CalibrationPlot} highlights the strong dependence on the statistical distributions on the sub-range and the packet window size.  Given that effective anomaly detection depends upon accurate models of the expected distribution it is worth considering the detection theory implications of these observations.

\section{Heavy-Tail Detection Theory}

Heavy-tail distributions are widely observed  \cite{nair2020fundamentals}, but only recently have these distributions become sufficiently precise that they can be used as background models for detection.  Heavy-tail distributions (like Zipf-Mandelbrot) differ significantly from their more commonly known light-tail counterparts (e.g., Gaussian, Poisson, ...), and have many surprising properties: high-likelihood of extreme events, divergent higher-order moments (e.g., infinite variance), and no central limit theorem.  In order to understand the implications of heavy-tail distributions on detection, a simple model is presented of the corresponding probability of detection ($p_{\rm d}$) and probability of false alarm ($p_{\rm fa}$) for these distributions.

\begin{figure}
\center{\includegraphics[width=0.85\columnwidth]{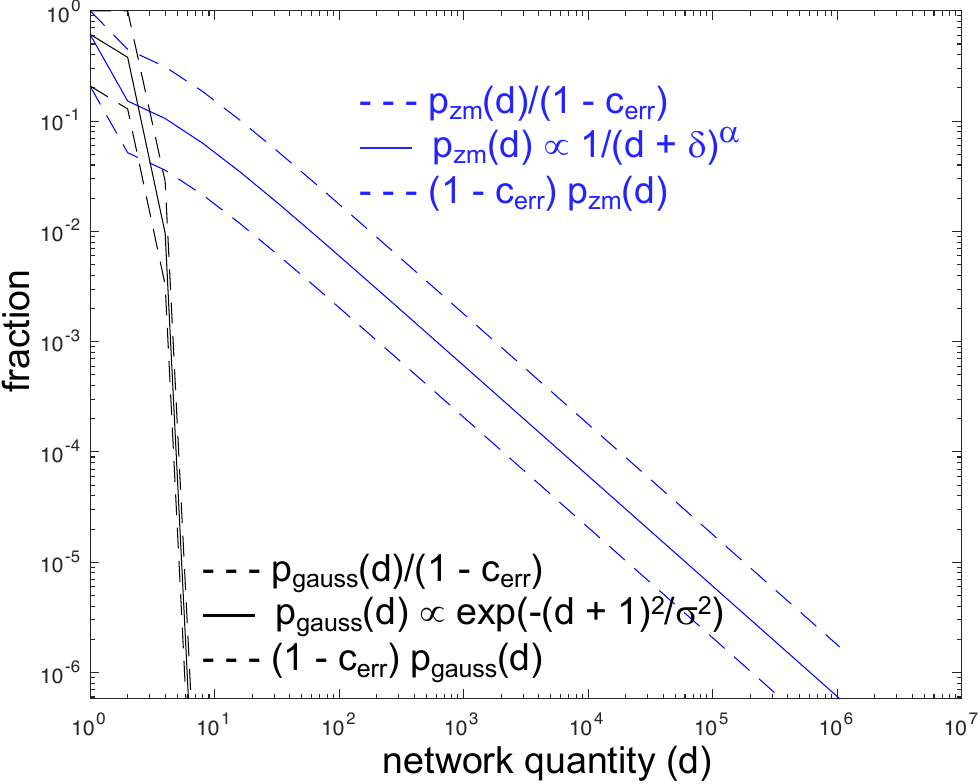}}
      	\caption{{\bf Heavy-Tail and Light-Tail Background Model.}  Heavy-tail distribution (solid blue line)  given by a Zipf-Mandelbrot distribution $p_{\rm zm}(d)$ with offset $\delta=0$ and exponent $\alpha=2$.    Light-tail distribution (solid black line) given by a Gaussian distribution $p_{\rm gauss}(d)$ with mean $\mu=1$ and variance $\sigma=0.43$.  Dashed lines show upper and lower error bounds on the model as determined by the parameter $c_{\rm err} = 2/3$.}
      	\label{fig:HeavyTailLightTail}
\end{figure}

Figure~\ref{fig:HeavyTailLightTail} shows representative heavy-tail and light-tail distributions that are each best-fits to underlying heavy-tail observations of the type observed in Figure~\ref{fig:CalibrationPlot} (clearly the light-tail distribution would be a poor fit).   The simple detection theory model begins by specifying upper and lower bounds of the models within which all background and  target observations are expected to be found.  These ranges are specified by the single parameter $c_{\rm err}$.  The bounds are defined in terms of multiplicative factors to represent equal logarithmic spacing which is more consistent with the high dynamic range observed in heavy-tail distributions.  Within this range a simple triangular model is used for the  expected distribution of background and target observations (see Appendix A).  These models can be used with a single threshold parameter $c_{\rm cut}$ to compute the expected probability of detection $p_{\rm d}(c_{\rm cut})$ and probability of false alarm $p_{\rm fa}(c_{\rm cut})$ to produce  the standard receiver operating characteristic (ROC) curves \cite{peterson1954theory, kerekes2008receiver} (see Appendix B).

\begin{figure}
\center{\includegraphics[width=0.85\columnwidth]{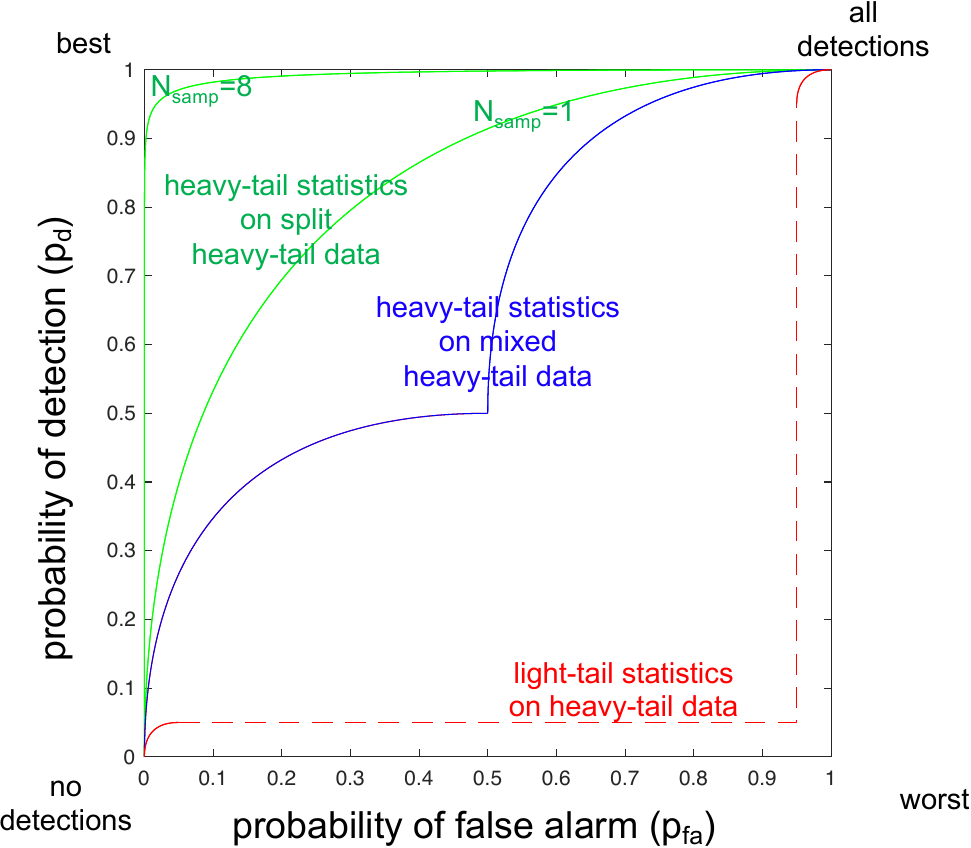}}
      	\caption{{\bf Model ROC Curve.}  Baseline model ($c_{\rm err} = 2/3$, $c_{\rm cut} = 1/3$, $d_{\rm max} = 2^{20}$) illustrating the performance of correct heavy-tail statistics  applied to correctly split heavy-tail data.  The performance can be further improved by requiring to detections appear in $N_{\rm samp} =8$ consecutive samples.  The performance is significantly degraded if a heavy-tail statistics is applied to mixed (unsplit) such that the underlying heavy-tail model is  correct  for only half of the bins.  The performance is very poor when light-tail statistics are used on heavy-tail data where it is likely only a single bin ${\rm f}_{d_{\rm max}}=0.05$ has the correct model.
}
      	\label{fig:ModelPdPfa}
\end{figure}
\begin{figure}
\center{\includegraphics[width=0.85\columnwidth]{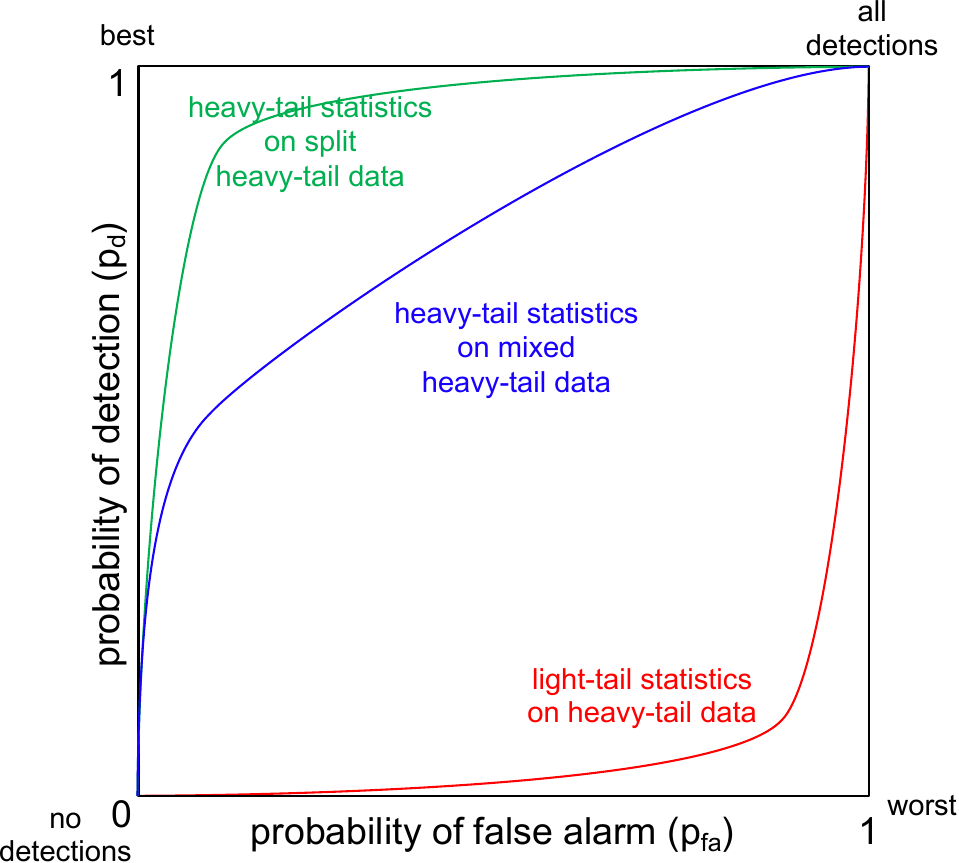}}
      	\caption{{\bf Notional ROC Curve.}   A  perhaps more accurate, but less precise version of Figure~\ref{fig:ModelPdPfa} highlighting that using the correct model on  correctly spit data provides good results (heavy-tail statistics with properly split heavy-tail data), using a  partially correct model degrades the performance (heavy-tail statistics with mixed heavy-tail data), and that using an entirely incorrect model results in very poor performance (light-tail statistics with heavy-tail data).
}
      	\label{fig:NotionalPdPfa}
\end{figure}

A simple baseline model covers the case when the background model and the observations are in agreement (see Figure~\ref{fig:ModelPdPfa}).  Given the frequency of heavy-tail observations (see Figure~\ref{fig:CalibrationPlot}) and that light-tail statistics are the more common analysis tool, it is important to explore the implications of using an incorrect model on the ROC curve.  Assuming the heavy-tail observations have a maximum value of $d_{\rm max}$ and are logarithmically binned into $\log_2(d_{\rm max})$ bins, a simple model of the effect of using a light-tail distribution on the heavy-tail data is to assume the light-tail distribution is correct for a single bin and incorrect for all others (Figure~\ref{fig:HeavyTailLightTail}).  Effectively, this means that the light-tail model will have a fractional accuracy of ${\rm f}_{d_{\rm max}} = 1/\log_2(d_{\rm max})$.  For a typical value of $d_{\rm max} \approx 2^{20}$, this implies a fractional accuracy of ${\rm f}_{d_{\rm max}} \approx 0.05$, which places the  ROC curve  in  either  the ``all detections'' or the ``no detections" regime.  The corresponding ROC curves for the all detections and no detections cases are shown in Figure~\ref{fig:ModelPdPfa}.

Figure~\ref{fig:CalibrationPlot} also highlights the importance of correctly sampling and splitting the data to compare with the correct heavy-tail distribution.  If the data is not sampled and split correctly, a heavy tail distribution may be observed with significantly different parameters than the appropriate background model.  A simple model for this effect can be extrapolated from the previous analysis.  Suppose using the incorrect heavy-tail model mischaracterizes half the bins, this is the same as setting ${\rm f}_{d_{\rm max}} \approx 0.5$ in the previous analysis, whose corresponding ROC curve is also shown in Figure~\ref{fig:ModelPdPfa}.

These simple models  provide precise estimates of ROC curves, but the real-world is far more complicated.  A  notional representation of these ROC curves, which is perhaps more accurate, but less precise is shown in Figure~\ref{fig:NotionalPdPfa}.  The overall interpretation of Figure~\ref{fig:NotionalPdPfa} is that using the correct model on the correctly split data provides good results (heavy-tail statistics with properly split heavy-tail data), using a  partially correct model degrades the performance (heavy-tail statistics with mixed heavy-tail data), and that using an entirely incorrect model results in very poor performance (light-tail statistics with heavy-tail data).

\section{Conclusions and Future Work}

Community-owned cyber space requires  community-based efforts to defend. A key to protecting our shared cyberspace are large-scale network observations that uphold the highest regard for privacy.  Careful sensor placement, focusing, and calibration with significant volumes of data are necessary for the deployment network sensors.  Novel focusing and calibration procedures using high-performance GraphBLAS anonymized hypersparse matrices are demonstrated on a multi-billion packet dataset.  Observed real-time processing rates are consistent with high-bandwidth links and show   significant data compression, confirming previously reported measurements.  The effectiveness of these procedures at focusing the traffic matrix is shown.  The underlying stable heavy-tail statistical distributions are also confirmed.  Effective anomaly detection depends upon accurate models of the expected distribution and the detection theory implications of these observations are explored with a simple model of the corresponding probability of detection ($p_{\rm d}$) and probability of false alarm ($p_{\rm fa}$).  This model highlights the criticality of network sensor focusing and calibration.  In practical terms,  once a sensor is properly focused and calibrated it can continuously observe  the network and minimize unbrokered network connections, which are two of the central tenets of good cybersecurity.

\section*{Acknowledgments}

The authors wish to acknowledge the following individuals for their contributions and support: Daniel Andersen, Sean Atkins, Chris Birardi, Bob Bond, Andy Bowne, Stephen Buckley, K Claffy, Cary Conrad, Chris Demchak, Alan Edelman, Garry Floyd, Jeff Gottschalk, Dhruv Gupta, Chris Hill, Kurt Keville, Charles Leiserson, Kirsten Malvey, Chase Milner, Sanjeev Mohindra,  Dave Martinez, Joseph McDonald, Heidi Perry, Christian Prothmann, Steve Rejto, Josh Rountree, Daniela Rus, Mark Sherman, Scott Weed, Adam Wierman, Marc Zissman.

\bibliographystyle{ieeetr}
\bibliography{FocusingCalibration}

\begin{thebibliography}{10}

\bibitem{atkins2021improvised}
S.~Atkins and C.~Lawson, ``An improvised patchwork: success and failure in
  cybersecurity policy for critical infrastructure,'' {\em Public
  Administration Review}, vol.~81, no.~5, pp.~847--861, 2021.

\bibitem{atkins2021cooperation}
S.~Atkins and C.~Lawson, ``Cooperation amidst competition: cybersecurity
  partnership in the us financial services sector,'' {\em Journal of
  Cybersecurity}, vol.~7, no.~1, 2021.

\bibitem{demchak2021achieving}
C.~Demchak, ``Achieving systemic resilience in a great systems conflict era,''
  {\em The Cyber Defense Review}, vol.~6, no.~2, pp.~51--70, 2021.

\bibitem{weed2022beyond}
S.~Weed, ``Beyond zero trust: Reclaiming blue cyberspace,'' Master's thesis,
  United States Army War College, 2022.

\bibitem{weed2023beyond}
S.~Atkins and C.~Lawson, ``Beyond zero trust: Reclaiming blue cyberspace with
  ai,'' {\em Cyber Defense Review}, vol.~7, no.~1, 2023.

\bibitem{kepner2021zero}
J.~Kepner, J.~Bernays, S.~Buckley, K.~Cho, C.~Conrad, L.~Daigle, K.~Erhardt,
  V.~Gadepally, B.~Greene, M.~Jones, R.~Knake, B.~Maggs, P.~Michaleas,
  C.~Meiners, A.~Morris, A.~Pentland, S.~Pisharody, S.~Powazek, A.~Prout,
  P.~Reiner, K.~Suzuki, K.~Takhashi, T.~Tauber, L.~Walker, and D.~Stetson,
  ``Zero botnets: An observe-pursue-counter approach.'' Belfer Center Reports,
  6 2021.

\bibitem{pisharody2021realizing}
S.~Pisharody, J.~Bernays, V.~Gadepally, M.~Jones, J.~Kepner, C.~Meiners,
  P.~Michaleas, A.~Tse, and D.~Stetson, ``Realizing forward defense in the
  cyber domain,'' in {\em 2021 IEEE High Performance Extreme Computing
  Conference (HPEC)}, pp.~1--7, IEEE, 2021.

\bibitem{pentland2022building}
A.~Pentland, ``Building a new economy: data, ai, and web3,'' {\em
  Communications of the ACM}, vol.~65, no.~12, pp.~27--29, 2022.

\bibitem{CAIDA2019}
``{\it CAIDA Anonymized Internet Traces Dataset (April 2008 - January 2019)}.''
  https://www.caida.org/catalog/datasets/passive\_dataset/.

\bibitem{CAIDA2022}
``{\it UCSD Network Telescope}.''
  https://www.caida.org/projects/network\_telescope/.

\bibitem{GCA2022}
``{\it Global Cyber Alliance}.'' https://www.globalcyberalliance.org/.

\bibitem{Greynoise2022}
``{\it Greynoise}.'' https://greynoise.io/.

\bibitem{MAWI2020}
``{\it MAWI Working Group Traffic Archive}.'' http://mawi.wide.ad.jp/mawi/.

\bibitem{Shadowserver2022}
``{\it Shadowserver Foundation}.'' https://www.shadowserver.org/.

\bibitem{kepner2020multi}
J.~Kepner, C.~Meiners, C.~Byun, S.~McGuire, T.~Davis, W.~Arcand, J.~Bernays,
  D.~Bestor, W.~Bergeron, V.~Gadepally, R.~Harnasch, M.~Hubbell, M.~Houle,
  M.~Jones, A.~Kirby, A.~Klein, L.~Milechin, J.~Mullen, A.~Prout, A.~Reuther,
  A.~Rosa, S.~Samsi, D.~Stetson, A.~Tse, C.~Yee, and P.~Michaleas,
  ``Multi-temporal analysis and scaling relations of 100,000,000,000 network
  packets,'' in {\em 2020 IEEE High Performance Extreme Computing Conference
  (HPEC)}, pp.~1--6, 2020.

\bibitem{bychkovskiy2003collaborative}
V.~Bychkovskiy, S.~Megerian, D.~Estrin, and M.~Potkonjak, ``A collaborative
  approach to in-place sensor calibration,'' in {\em Information Processing in
  Sensor Networks: Second International Workshop, IPSN 2003, Palo Alto, CA,
  USA, April 22--23, 2003 Proceedings}, pp.~301--316, Springer, 2003.

\bibitem{Cisco2017}
``{\it Cisco Visual Networking Index: Forecast and Trends}.''
  https://newsroom.cisco.com/press-release-content?articleId=1955935.

\bibitem{Cisco2018-2023}
``{\it Cisco Visual Networking Index: Forecast and Trends, 2018–2023}.''
  https://www.cisco.com/c/en/us/solutions/collateral/executive-perspectives/annual-internet-report/white-paper-c11-741490.html.

\bibitem{kepner16mathematical}
J.~{Kepner}, P.~{Aaltonen}, D.~{Bader}, A.~{Bulu{\c{c}}}, F.~{Franchetti},
  J.~{Gilbert}, D.~{Hutchison}, M.~{Kumar}, A.~{Lumsdaine}, H.~{Meyerhenke},
  S.~{McMillan}, C.~{Yang}, J.~D. {Owens}, M.~{Zalewski}, T.~{Mattson}, and
  J.~{Moreira}, ``Mathematical foundations of the graphblas,'' in {\em 2016
  IEEE High Performance Extreme Computing Conference (HPEC)}, pp.~1--9, 2016.

\bibitem{buluc17design}
A.~{Bulu{\c{c}}}, T.~{Mattson}, S.~{McMillan}, J.~{Moreira}, and C.~{Yang},
  ``Design of the graphblas api for c,'' in {\em 2017 IEEE International
  Parallel and Distributed Processing Symposium Workshops (IPDPSW)},
  pp.~643--652, 2017.

\bibitem{yang2018implementing}
C.~Yang, A.~Bulu{\c{c}}, and J.~D. Owens, ``Implementing push-pull efficiently
  in graphblas,'' in {\em Proceedings of the 47th International Conference on
  Parallel Processing}, pp.~1--11, 2018.

\bibitem{kepner2018mathematics}
J.~Kepner and H.~Jananthan, {\em Mathematics of big data: Spreadsheets,
  databases, matrices, and graphs}.
\newblock MIT Press, 2018.

\bibitem{davis2019algorithm}
T.~A. Davis, ``Algorithm 1000: Suitesparse: Graphblas: Graph algorithms in the
  language of sparse linear algebra,'' {\em ACM Transactions on Mathematical
  Software (TOMS)}, vol.~45, no.~4, pp.~1--25, 2019.

\bibitem{mattson2019lagraph}
T.~Mattson, T.~A. Davis, M.~Kumar, A.~Buluc, S.~McMillan, J.~Moreira, and
  C.~Yang, ``Lagraph: A community effort to collect graph algorithms built on
  top of the graphblas,'' in {\em 2019 IEEE International Parallel and
  Distributed Processing Symposium Workshops (IPDPSW)}, pp.~276--284, IEEE,
  2019.

\bibitem{cailliau2019redisgraph}
P.~Cailliau, T.~Davis, V.~Gadepally, J.~Kepner, R.~Lipman, J.~Lovitz, and
  K.~Ouaknine, ``Redisgraph graphblas enabled graph database,'' in {\em 2019
  IEEE International Parallel and Distributed Processing Symposium Workshops
  (IPDPSW)}, pp.~285--286, IEEE, 2019.

\bibitem{davis2019write}
T.~A. Davis, M.~Aznaveh, and S.~Kolodziej, ``Write quick, run fast: Sparse deep
  neural network in 20 minutes of development time via suitesparse:
  Graphblas,'' in {\em 2019 IEEE High Performance Extreme Computing Conference
  (HPEC)}, pp.~1--6, IEEE, 2019.

\bibitem{aznaveh2020parallel}
M.~Aznaveh, J.~Chen, T.~A. Davis, B.~Hegyi, S.~P. Kolodziej, T.~G. Mattson, and
  G.~Sz{\'a}rnyas, ``Parallel graphblas with openmp,'' in {\em 2020 Proceedings
  of the SIAM Workshop on Combinatorial Scientific Computing}, pp.~138--148,
  SIAM, 2020.

\bibitem{brock2021introduction}
B.~Brock, A.~Bulu{\c{c}}, T.~G. Mattson, S.~McMillan, and J.~E. Moreira,
  ``Introduction to graphblas 2.0,'' in {\em 2021 IEEE International Parallel
  and Distributed Processing Symposium Workshops (IPDPSW)}, pp.~253--262, IEEE,
  2021.

\bibitem{pelletier2021graphblas}
M.~Pelletier, W.~Kimmerer, T.~A. Davis, and T.~G. Mattson, ``The graphblas in
  julia and python: the pagerank and triangle centralities,'' in {\em 2021 IEEE
  High Performance Extreme Computing Conference (HPEC)}, pp.~1--7, 2021.

\bibitem{jones2022graphblas}
M.~Jones, J.~Kepner, D.~Andersen, A.~Buluç, C.~Byun, K.~Claffy, T.~Davis,
  W.~Arcand, J.~Bernays, D.~Bestor, W.~Bergeron, V.~Gadepally, M.~Houle,
  M.~Hubbell, H.~Jananthan, A.~Klein, C.~Meiners, L.~Milechin, J.~Mullen,
  S.~Pisharody, A.~Prout, A.~Reuther, A.~Rosa, S.~Samsi, J.~Sreekanth,
  D.~Stetson, C.~Yee, and P.~Michaleas, ``Graphblas on the edge: Anonymized
  high performance streaming of network traffic,'' in {\em 2022 IEEE High
  Performance Extreme Computing Conference (HPEC)}, pp.~1--8, 2022.

\bibitem{trigg2022hypersparse}
T.~Trigg, C.~Meiners, S.~Pisharody, H.~Jananthan, M.~Jones, A.~Michaleas,
  T.~Davis, E.~Welch, W.~Arcand, D.~Bestor, W.~Bergeron, C.~Byun, V.~Gadepally,
  M.~Houle, M.~Hubbell, A.~Klein, P.~Michaleas, L.~Milechin, J.~Mullen,
  A.~Prout, A.~Reuther, A.~Rosa, S.~Samsi, D.~Stetson, C.~Yee, and J.~Kepner,
  ``Hypersparse network flow analysis of packets with graphblas,'' in {\em 2022
  IEEE High Performance Extreme Computing Conference (HPEC)}, pp.~1--7, 2022.

\bibitem{kepner2021vertical}
J.~Kepner, T.~Davis, C.~Byun, W.~Arcand, D.~Bestor, W.~Bergeron, V.~Gadepally,
  M.~Houle, M.~Hubbell, M.~Jones, A.~Klein, L.~Milechin, J.~Mullen, A.~Prout,
  A.~Reuther, A.~Rosa, S.~Samsi, C.~Yee, and P.~Michaleas, ``Vertical,
  temporal, and horizontal scaling of hierarchical hypersparse graphblas
  matrices,'' in {\em 2021 IEEE High Performance Extreme Computing Conference
  (HPEC)}, pp.~1--6, 2021.

\bibitem{kepner2021spatial}
J.~Kepner, M.~Jones, D.~Andersen, A.~Buluç, C.~Byun, K.~Claffy, T.~Davis,
  W.~Arcand, J.~Bernays, D.~Bestor, W.~Bergeron, V.~Gadepally, M.~Houle,
  M.~Hubbell, A.~Klein, C.~Meiners, L.~Milechin, J.~Mullen, S.~Pisharody,
  A.~Prout, A.~Reuther, A.~Rosa, S.~Samsi, D.~Stetson, A.~Tse, C.~Yee, and
  P.~Michaleas, ``Spatial temporal analysis of 40,000,000,000,000 internet
  darkspace packets,'' in {\em 2021 IEEE High Performance Extreme Computing
  Conference (HPEC)}, pp.~1--8, 2021.

\bibitem{tumeo2010efficient}
A.~Tumeo, O.~Villa, and D.~Sciuto, ``Efficient pattern matching on gpus for
  intrusion detection systems,'' in {\em Proceedings of the 7th ACM
  International Conference on Computing Frontiers}, CF '10, (New York, NY,
  USA), p.~87–88, Association for Computing Machinery, 2010.

\bibitem{kumar2018ibm}
M.~Kumar, W.~P. Horn, J.~Kepner, J.~E. Moreira, and P.~Pattnaik, ``Ibm power9
  and cognitive computing,'' {\em IBM Journal of Research and Development},
  vol.~62, no.~4/5, pp.~10--1, 2018.

\bibitem{ezick2019combining}
J.~Ezick, T.~Henretty, M.~Baskaran, R.~Lethin, J.~Feo, T.-C. Tuan, C.~Coley,
  L.~Leonard, R.~Agrawal, B.~Parsons, and W.~Glodek, ``Combining tensor
  decompositions and graph analytics to provide cyber situational awareness at
  hpc scale,'' in {\em 2019 IEEE High Performance Extreme Computing Conference
  (HPEC)}, pp.~1--7, 2019.

\bibitem{gera2020traversing}
P.~Gera, H.~Kim, P.~Sao, H.~Kim, and D.~Bader, ``Traversing large graphs on
  gpus with unified memory,'' {\em Proceedings of the VLDB Endowment}, vol.~13,
  no.~7, pp.~1119--1133, 2020.

\bibitem{azad2020evaluation}
A.~Azad, M.~M. Aznaveh, S.~Beamer, M.~Blanco, J.~Chen, L.~D'Alessandro,
  R.~Dathathri, T.~Davis, K.~Deweese, J.~Firoz, H.~A. Gabb, G.~Gill, B.~Hegyi,
  S.~Kolodziej, T.~M. Low, A.~Lumsdaine, T.~Manlaibaatar, T.~G. Mattson,
  S.~McMillan, R.~Peri, K.~Pingali, U.~Sridhar, G.~Szarnyas, Y.~Zhang, and
  Y.~Zhang, ``Evaluation of graph analytics frameworks using the gap benchmark
  suite,'' in {\em 2020 IEEE International Symposium on Workload
  Characterization (IISWC)}, pp.~216--227, 2020.

\bibitem{du2021interactive}
Z.~Du, O.~A. Rodriguez, J.~Patchett, and D.~A. Bader, ``Interactive graph
  stream analytics in arkouda,'' {\em Algorithms}, vol.~14, no.~8, p.~221,
  2021.

\bibitem{acer2021exagraph}
S.~Acer, A.~Azad, E.~G. Boman, A.~Buluç, K.~D. Devine, S.~Ferdous, N.~Gawande,
  S.~Ghosh, M.~Halappanavar, A.~Kalyanaraman, A.~Khan, M.~Minutoli, A.~Pothen,
  S.~Rajamanickam, O.~Selvitopi, N.~R. Tallent, and A.~Tumeo, ``Exagraph: Graph
  and combinatorial methods for enabling exascale applications,'' {\em The
  International Journal of High Performance Computing Applications}, vol.~35,
  no.~6, pp.~553--571, 2021.

\bibitem{blanco2021delayed}
M.~P. Blanco, S.~McMillan, and T.~M. Low, ``Delayed asynchronous iterative
  graph algorithms,'' in {\em 2021 IEEE High Performance Extreme Computing
  Conference (HPEC)}, pp.~1--7, IEEE, 2021.

\bibitem{ahmed2021online}
N.~K. Ahmed, N.~Duffield, and R.~A. Rossi, ``Online sampling of temporal
  networks,'' {\em ACM Transactions on Knowledge Discovery from Data (TKDD)},
  vol.~15, no.~4, pp.~1--27, 2021.

\bibitem{azad2021combinatorial}
A.~Azad, O.~Selvitopi, M.~T. Hussain, J.~R. Gilbert, and A.~Bulu{\c{c}},
  ``Combinatorial blas 2.0: Scaling combinatorial algorithms on
  distributed-memory systems,'' {\em IEEE Transactions on Parallel and
  Distributed Systems}, vol.~33, no.~4, pp.~989--1001, 2021.

\bibitem{koutra2021power}
D.~Koutra, ``The power of summarization in graph mining and learning: smaller
  data, faster methods, more interpretability,'' {\em Proceedings of the VLDB
  Endowment}, vol.~14, no.~13, pp.~3416--3416, 2021.

\bibitem{hofstede2014flow}
R.~Hofstede, P.~{\v{C}}eleda, B.~Trammell, I.~Drago, R.~Sadre, A.~Sperotto, and
  A.~Pras, ``Flow monitoring explained: From packet capture to data analysis
  with netflow and ipfix,'' {\em IEEE Communications Surveys \& Tutorials},
  vol.~16, no.~4, pp.~2037--2064, 2014.

\bibitem{sommer2003bro}
R.~Sommer, ``Bro: An open source network intrusion detection system,'' {\em
  Security, E-learning, E-Services, 17. DFN-Arbeitstagung {\"u}ber
  Kommunikationsnetze}, 2003.

\bibitem{lucente2008pmacct}
P.~Lucente, ``pmacct: steps forward interface counters,'' {\em Tech. Rep.},
  2008.

\bibitem{fan2004prefix}
J.~Fan, J.~Xu, M.~H. Ammar, and S.~B. Moon, ``Prefix-preserving ip address
  anonymization: measurement-based security evaluation and a new
  cryptography-based scheme,'' {\em Computer Networks}, vol.~46, no.~2,
  pp.~253--272, 2004.

\bibitem{kepner19hypersparse}
J.~{Kepner}, K.~{Cho}, K.~{Claffy}, V.~{Gadepally}, P.~{Michaleas}, and
  L.~{Milechin}, ``Hypersparse neural network analysis of large-scale internet
  traffic,'' in {\em 2019 IEEE High Performance Extreme Computing Conference
  (HPEC)}, pp.~1--11, 2019.

\bibitem{karvanen2003measuring}
J.~Karvanen and A.~Cichocki, ``Measuring sparseness of noisy signals,'' in {\em
  4th International Symposium on Independent Component Analysis and Blind
  Signal Separation}, pp.~125--130, 2003.

\bibitem{soule2004identify}
A.~Soule, A.~Nucci, R.~Cruz, E.~Leonardi, and N.~Taft, ``How to identify and
  estimate the largest traffic matrix elements in a dynamic environment,'' in
  {\em ACM SIGMETRICS Performance Evaluation Review}, vol.~32, pp.~73--84, ACM,
  2004.

\bibitem{zhang2005estimating}
Y.~Zhang, M.~Roughan, C.~Lund, and D.~L. Donoho, ``Estimating point-to-point
  and point-to-multipoint traffic matrices: an information-theoretic
  approach,'' {\em IEEE/ACM Transactions on Networking (TON)}, vol.~13, no.~5,
  pp.~947--960, 2005.

\bibitem{mucha2010community}
P.~J. Mucha, T.~Richardson, K.~Macon, M.~A. Porter, and J.-P. Onnela,
  ``Community structure in time-dependent, multiscale, and multiplex
  networks,'' {\em science}, vol.~328, no.~5980, pp.~876--878, 2010.

\bibitem{tune2013internet}
P.~Tune, M.~Roughan, H.~Haddadi, and O.~Bonaventure, ``Internet traffic
  matrices: A primer,'' {\em Recent Advances in Networking}, vol.~1, pp.~1--56,
  2013.

\bibitem{kepner19streaming}
J.~{Kepner}, V.~{Gadepally}, L.~{Milechin}, S.~{Samsi}, W.~{Arcand},
  D.~{Bestor}, W.~{Bergeron}, C.~{Byun}, M.~{Hubbell}, M.~{Houle}, M.~{Jones},
  A.~{Klein}, P.~{Michaleas}, J.~{Mullen}, A.~{Prout}, A.~{Rosa}, C.~{Yee}, and
  A.~{Reuther}, ``Streaming 1.9 billion hypersparse network updates per second
  with d4m,'' in {\em 2019 IEEE High Performance Extreme Computing Conference
  (HPEC)}, pp.~1--6, 2019.

\bibitem{nair2020fundamentals}
J.~Nair, A.~Wierman, and B.~Zwart, ``The fundamentals of heavy tails:
  Properties, emergence, and estimation,'' {\em Preprint, California Institute
  of Technology}, 2020.

\bibitem{kepner2022new}
J.~Kepner, K.~Cho, K.~Claffy, V.~Gadepally, S.~McGuire, L.~Milechin, W.~Arcand,
  D.~Bestor, W.~Bergeron, C.~Byun, M.~Hubbell, M.~Houle, M.~Jones, A.~Prout,
  A.~Reuther, A.~Rosa, S.~Samsi, C.~Yee, and P.~Michaleas, ``New phenomena in
  large-scale internet traffic,'' in {\em Massive Graph Analytics} (D.~Bader,
  ed.), pp.~1--53, Chapman and Hall/CRC, 2022.

\bibitem{bulucc2009parallel}
A.~Bulu{\c{c}}, J.~T. Fineman, M.~Frigo, J.~R. Gilbert, and C.~E. Leiserson,
  ``Parallel sparse matrix-vector and matrix-transpose-vector multiplication
  using compressed sparse blocks,'' in {\em Proceedings of the twenty-first
  annual symposium on Parallelism in algorithms and architectures},
  pp.~233--244, 2009.

\bibitem{kepner202075}
J.~Kepner, T.~Davis, C.~Byun, W.~Arcand, D.~Bestor, W.~Bergeron, V.~Gadepally,
  M.~Hubbell, M.~Houle, M.~Jones, A.~Klein, P.~Michaleas, L.~Milechin,
  J.~Mullen, A.~Prout, A.~Rosa, S.~Samsi, C.~Yee, and A.~Reuther,
  ``75,000,000,000 streaming inserts/second using hierarchical hypersparse
  graphblas matrices,'' in {\em 2020 IEEE International Parallel and
  Distributed Processing Symposium Workshops (IPDPSW)}, pp.~207--210, 2020.

\bibitem{collet2018zstandard}
Y.~Collet and M.~Kucherawy, ``Zstandard compression and the application/zstd
  media type,'' tech. rep., 2018.

\bibitem{arsenault2021hybrid}
B.~Arsenault, ``Hybrid workforce security,'' {\em Microsoft Digital Defense
  Report}, pp.~89--108, 2021.

\bibitem{peterson1954theory}
W.~Peterson, T.~Birdsall, and W.~Fox, ``The theory of signal detectability,''
  {\em Transactions of the IRE professional group on information theory},
  vol.~4, no.~4, pp.~171--212, 1954.

\bibitem{kerekes2008receiver}
J.~Kerekes, ``Receiver operating characteristic curve confidence intervals and
  regions,'' {\em IEEE Geoscience and Remote Sensing Letters}, vol.~5, no.~2,
  pp.~251--255, 2008.

\end{thebibliography}

\section*{Appendix A: Background and Target Distributions}

Simple triangular models of the expected distributions around the background model can be created with just two parameters: $c_{\rm err}$ and $c_{\rm cut}$.  Let $x$ be observed distance from the expected background distribution $p(d)$, where $d$ is a measured network quantity (see Table~\ref{tab:Aggregates}). $c_{\rm err}$ sets the minimum and maximum values of the domains of these distributions.  
\begin{eqnarray*}
0 < &c_{\rm err}& < 1 \\
x_{\rm min} &=& (1 - c_{\rm err}) \\
x_{\rm max} &=& 1/(1 - c_{\rm err})
\end{eqnarray*}
where $x_{\rm min} < x < x_{\rm max}$.  $c_{\rm cut}$ sets the minimum and maximum cut values used within the cumulative distributions to label background and targets
\begin{eqnarray*}
0 < &c_{\rm cut}& < c_{\rm err} \\
x_{\rm min}^{\rm cut} &=& (1 - c_{\rm cut}) > x_{\rm min} \\
x_{\rm max}^{\rm cut} &=& 1/(1 - c_{\rm cut}) < x_{\rm max} 
\end{eqnarray*}
For a given value of $c_{\rm cut}$, observations where $x_{\rm min}^{\rm cut} < x < x_{\rm max}^{\rm cut}$ are declared as background and observations where $x <x_{\rm min}^{\rm cut}$ or  $x > x_{\rm max}^{\rm cut}$ are declared as targets.  From these definitions, the relative, normalized, and cumulative lower/higher background and target models can be computed.  

\begin{figure}
\center{\includegraphics[width=0.85\columnwidth]{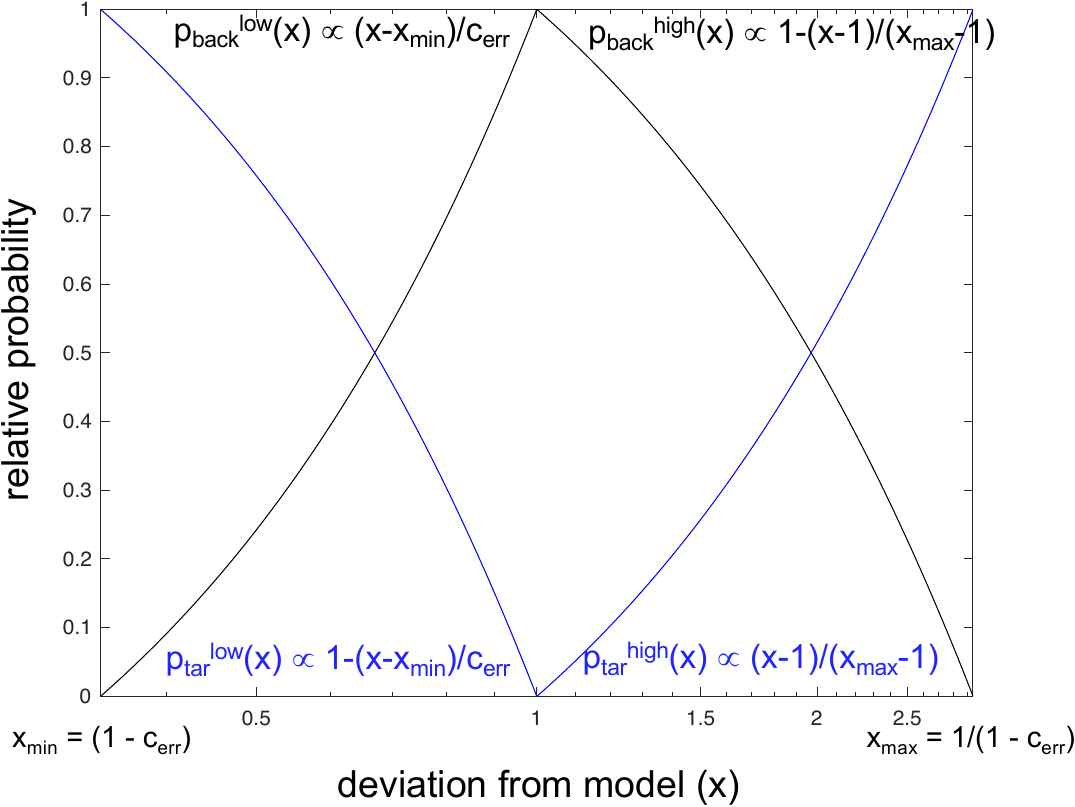}}
      	\caption{{\bf Relative Background and Target Probabilities.}  The expected distribution of background and targets as a function of the distance $x$ from the background model $p(d)$ for $c_{\rm err} = 2/3$. $p_{\rm back/tar}^{low/high}(x)$ is the probability of a background/target being lower/higher that the model.  Background is expected to peak around the background model while targets are more likely away from the background model.}
      	\label{fig:RelativeTarBack}
\end{figure}

\begin{figure}
\center{\includegraphics[width=0.85\columnwidth]{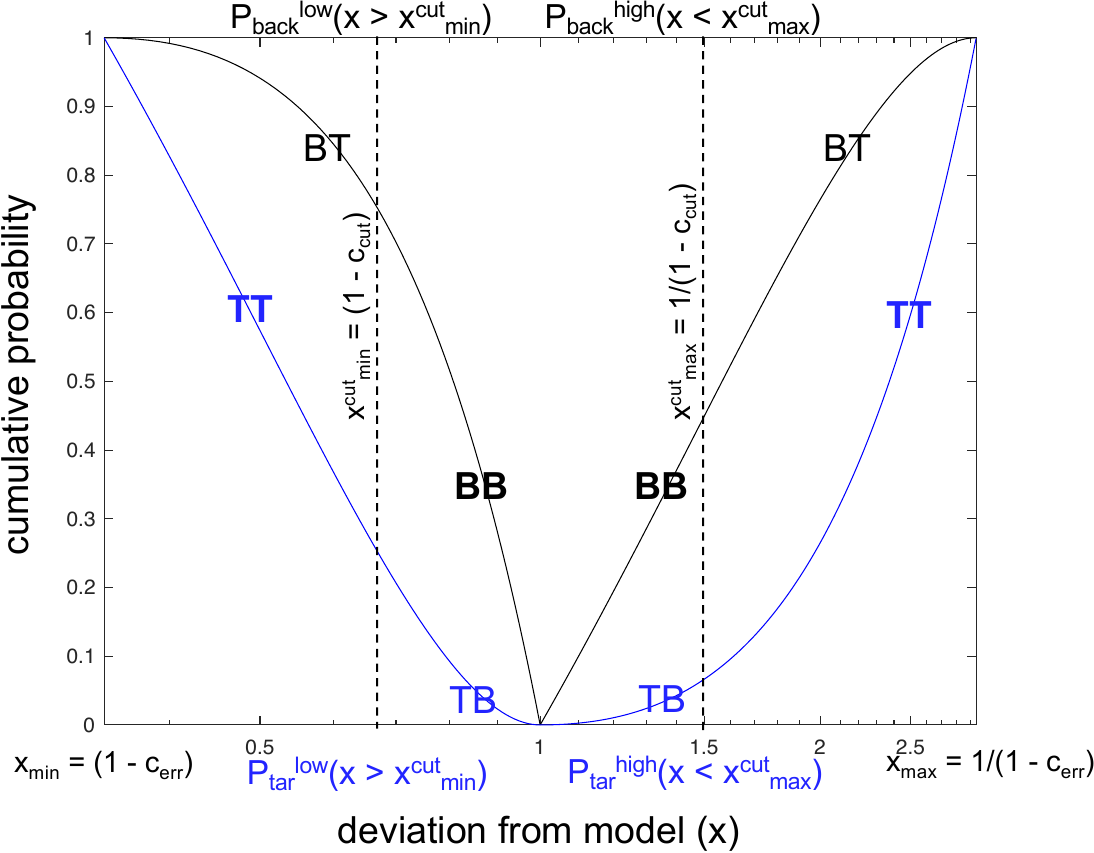}}
      	\caption{{\bf Cumulative Background and Target Probabilities.}  The expected probability of background and targets relative to $x^{\rm cut}$ threshold for $c_{\rm err} = 2/3$ and $c_{\rm cut} = 1/3$. $P_{\rm back/tar}^{low/high}(x ~ {>}/{<} ~ x_{\rm min/max}^{\rm cut})$ is the probability of a background/target being lower/higher than the corresponding cut threshold $x_{\rm min/max}^{\rm cut}$.  From these probabilities the four outcomes of labeling can be computed: true targets labeled as targets (\textbf{\textsf{\color{blue} TT}}), true targets labeled as background ({\textsf{\color{blue} TB}}), true background labeled as target ({\textsf{BT}}), and true background labeled as background (\textbf{\textsf{BB}}).}
      	\label{fig:CumulativeTarBack}
\end{figure}

\noindent (1) Relative, normalized, cumulative lower background model
\begin{eqnarray*}
  p_{\rm back}^{\rm low}(x) &\propto& (x - x_{\rm min})/c_{\rm err} = 1 + (x-1)/c_{\rm err} \\
  p_{\rm back}^{\rm low}(x) &=& 2(x - (1 - c_{\rm err}))/c_{\rm err}^2  \\
  P_{\rm back}^{\rm low}(x) &=& (1-x)(2 c_{\rm err} + x -1)/c_{\rm err}^2
\end{eqnarray*}
(2) Relative, normalized, cumulative higher background model
\begin{eqnarray*}
  p_{\rm back}^{\rm high}(x) &\propto& 1-(x-1)/(x_{\rm max}-1)  = x + (1-x)/c_{\rm err} \\
  p_{\rm back}^{\rm high}(x) &=& (2/c_{\rm err})(1 - c_{\rm err})(x + (1-x)/c_{\rm err})  \\
  P_{\rm back}^{\rm high}(x) &=& (1-c_{\rm err})(x-1)(c_{\rm err} x + c_{\rm err} - x +1)/c_{\rm err}^2
\end{eqnarray*}
(3) Relative, normalized, cumulative lower target model
\begin{eqnarray*}
  p_{\rm tar}^{\rm low}(x) &\propto& 1-(x-x_{\rm min})/c_{\rm err}  = (1-x)/c_{\rm err} \\
  p_{\rm tar}^{\rm low}(x) &=& 2(1 - x)/c_{\rm err}^2   \\
  P_{\rm tar}^{\rm low}(x) &=& (x-1)^2/c_{\rm err}^2
\end{eqnarray*}
(4) Relative, normalized, cumulative higher target model
\begin{eqnarray*}
  p_{\rm tar}^{\rm high}(x) &\propto& (x-1)/(x_{\rm max}-1)   = (1- c_{\rm err})(x-1)/c_{\rm err} \\
  p_{\rm tar}^{\rm high}(x) &=& (2/c_{\rm err}^2)(1 - c_{\rm err})^2(x - 1)   \\
  P_{\rm tar}^{\rm high}(x) &=& (c_{\rm err} - 1)^2 (x - 1)^2/c_{\rm err}^2
\end{eqnarray*}
Figure~\ref{fig:RelativeTarBack} and Figure~\ref{fig:CumulativeTarBack} shows the above distributions for the specific values of $c_{\rm err} = 2/3$ and $c_{\rm cut} = 1/3$.

\section*{Appendix B: Background and Target Labeling}

The four outcomes of labeling targets and background can be computed from the corresponding cumulative probabilities $P_{\rm back/tar}^{low/high}(x ~ {>}/{<} ~ x_{\rm min/max}^{\rm cut})$.  The definitions of the outcomes and their corresponding formula are \cite{kerekes2008receiver}
\begin{itemize}
\item [\textbf{\textsf{\color{blue} TT}}] True target labeled as a target (correct detection)
  $$
     p(\textbf{\textsf{\color{blue} TT}}) = (1 - P_{\rm tar}^{\rm low}(x > x^{\rm cut}_{\rm min}) + 1 - P_{\rm tar}^{\rm high}(x < x^{\rm cut}_{\rm max}))/2
  $$
\item [\textsf{\color{blue} TB}] True target labeled as background (missed detection)
  $$
      p(\textsf{\color{blue} TB}) = (P_{\rm tar}^{\rm low}(x > x^{\rm cut}_{\rm min}) + P_{\rm tar}^{\rm high}(x < x^{\rm cut}_{\rm max}))/2
  $$
\item [\textsf{BT}] True background labeled as target (false alarm)
  $$
      p(\textsf{BT}) = (1- P_{\rm back}^{\rm low}(x > x^{\rm cut}_{\rm min}) + 1- P_{\rm back}^{\rm high}(x < x^{\rm cut}_{\rm max}))/2
  $$
\item [\textbf{\textsf{BB}}] True background labeled as background (correct nondetection)
  $$
     p(\textbf{\textsf{BB}}) = (P_{\rm back}^{\rm low}(x > x^{\rm cut}_{\rm min}) + P_{\rm back}^{\rm high}(x < x^{\rm cut}_{\rm max}))/2
  $$
\end{itemize}
From these definitions, baseline formulas for the probability of detection ($p_{\rm det}$) and probability of false alarm ($p_{\rm fa}$) can be derived that are ultimately a function of  the single parameter $c_{\rm cut}$ via $x^{\rm cut}_{\rm min/max}$
\begin{eqnarray*}
p_{\rm det}(c_{\rm cut}) &=& \frac{N(\textbf{\textsf{\color{blue} TT}})}{N(\textbf{\textsf{\color{blue} TT}}) + N(\textsf{\color{blue} TB})} = p(\textbf{\textsf{\color{blue} TT}}) \\
p_{\rm fa}(c_{\rm cut}) &=& \frac{N(\textsf{BT})}{N(\textsf{BT}) + N(\textbf{\textsf{BB}})} = p(\textsf{BT})
\end{eqnarray*}
where $N()$ is the count of the corresponding outcome.  Plotting $p_{\rm det}(c_{\rm cut})$ versus $p_{\rm fa}(c_{\rm cut})$ for different values of $c_{\rm cut}$ produces the standard receiver operating characteristic (ROC) curve \cite{peterson1954theory} (see Figure~\ref{fig:ModelPdPfa}).

A standard enhancement to this model is to integrate over time and assert that true detections are coherent over $N_{\rm samp}$ consecutive samples.  This does not change $p_{\rm det}$ but can have a significant effect on $p_{\rm fa}$. Assuming the background is random and observations are binomially distributed between $\textbf{\textsf{BB}}$ and $\textsf{BT}$, the probability that a true background will be randomly categorized as a detection in $k$ out of $N_{\rm samp}$ is given by
$$
  p(n(\textbf{\textsf{BB}}) = k) = \binom{N_{\rm samp}}{k} p(\textbf{\textsf{BB}})^k (1 - p(\textbf{\textsf{BB}})^{N_{\rm samp} -k}
$$
If an observation must appear as a detection in all $N_{\rm samp}$ to be a true detection, then a false alarm requires $k=0$ resulting in 
\begin{eqnarray*}
  p_{\rm fa}(c_{\rm cut}) &=& p(n(\textbf{\textsf{BB}}) = 0) \\
                                      &=& (1 - p(\textbf{\textsf{BB}}))^{N_{\rm samp}} \\
                                      &=& p(\textsf{BT})^{N_{\rm samp}}
\end{eqnarray*}
since $ p(\textsf{BT}) = 1 - p(\textbf{\textsf{BB}})$.   This has a significant positive impact on the shape of the ROC curve (see Figure~\ref{fig:ModelPdPfa}).

The baseline model covers the case when the background model and the observations are in agreement.  Given the frequency of heavy-tail observations (see Figure~\ref{fig:CalibrationPlot}) and that light-tail statistics are the more common analysis tool, it is important to explore the ROC curve implications of using a light-tail model on heavy-tail observations.  Assuming the heavy-tail observations have a maximum value of $d_{\rm max}$ and are logarithmically binned into $\log_2(d_{\rm max})$ bins, a simple model of the effect of using a light-tail distribution on the heavy-tail data is to assume the light-tail distribution is correct for a single bin and incorrect for all others (Figure~\ref{fig:HeavyTailLightTail}).  Effectively, this means that the light-tail model will have a fractional accuracy of ${\rm f}_{d_{\rm max}} = 1/\log_2(d_{\rm max})$.  For a typical value of $d_{\rm max} \approx 2^{20}$, this implies a fractional accuracy of ${\rm f}_{d_{\rm max}} \approx 0.05$, which places the  ROC curve  in  either  the ``all detections'' or the ``no detections" regime.  The all detections regime can be estimated by declaring all observations not in the correctly modeled bin as detections because they satisfy the $x < x^{\rm cut}_{\rm min}$ or $x > x^{\rm cut}_{\rm max}$ criteria. Only ${\rm f}_{d_{\rm max}}$ of the observations could possibly be categorized with the model and the remaining $1-{\rm f}_{d_{\rm max}}$ will be declared detections
\begin{eqnarray*}
p_{\rm det}(c_{\rm cut}) &=& p(\textbf{\textsf{\color{blue} TT}}) {\rm f}_{d_{\rm max}} + 1-{\rm f}_{d_{\rm max}} \\
p_{\rm fa}(c_{\rm cut}) &=& p(\textsf{BT}) {\rm f}_{d_{\rm max}} + 1-{\rm f}_{d_{\rm max}}
\end{eqnarray*}
Likewise, the no detections regime can be estimated  by declaring all observations not in the correct bin as non-detections because they lie outside the domain of the model the $x_{\rm min} < x < x_{\rm max}$.  Only ${\rm f}_{d_{\rm max}}$ of the observations could possibly be categorized with the model and the remaining $1-{\rm f}_{d_{\rm max}}$ will be declared non-detections 
\begin{eqnarray*}
p_{\rm det}(c_{\rm cut}) &=& p(\textbf{\textsf{\color{blue} TT}}) {\rm f}_{d_{\rm max}} \\
p_{\rm fa}(c_{\rm cut}) &=& p(\textsf{BT}) {\rm f}_{d_{\rm max}}
\end{eqnarray*}
The corresponding ROC curves for the all detections and no detections cases are shown in Figure~\ref{fig:ModelPdPfa}.

\end{document}